\documentclass[aps,prl,twocolumn,superscriptaddress]{revtex4}
\usepackage{graphicx}

\newcommand{\eq}{\begin{equation}}
\newcommand{\fine}{\end{equation}}
\newcommand{\ket}[1]{|#1\rangle}
\newcommand{\bra}[1]{\langle #1|}

\begin{document}
\title{Polarization control of single photon quantum orbital angular momentum states}
\author{E. Nagali}
\author{F. Sciarrino}
\email{fabio.sciarrino@uniroma1.it} \affiliation{Dipartimento di
Fisica dell'Universit\`{a} ``La Sapienza'' and Consorzio Nazionale
Interuniversitario per le Scienze Fisiche della Materia, Roma 00185,
Italy}
\author{F. De Martini}
\affiliation{Dipartimento di Fisica dell'Universit\`{a} ``La
Sapienza'' and Consorzio Nazionale Interuniversitario per le Scienze
Fisiche della Materia, Roma 00185, Italy}
\affiliation{Accademia Nazionale dei Lincei, via della Lungara 10,
Roma 00165, Italy}
\author{B. Piccirillo}
\affiliation{Dipartimento di Scienze Fisiche, Universit\`{a} di
Napoli ``Federico II'', Compl.\ Univ.\ di Monte S. Angelo, 80126
Napoli, Italy}
\affiliation{Consorzio Nazionale Interuniversitario per le Scienze
Fisiche della Materia, Napoli}
\author{E. Karimi}
\author{L. Marrucci}
\affiliation{Dipartimento di Scienze Fisiche, Universit\`{a} di
Napoli ``Federico II'', Compl.\ Univ.\ di Monte S. Angelo, 80126
Napoli, Italy}
\affiliation{ CNR-INFM Coherentia, Compl.\ Univ.\ di
Monte S. Angelo, 80126 Napoli, Italy}
\author{E. Santamato}
\affiliation{Dipartimento di Scienze Fisiche, Universit\`{a} di
Napoli ``Federico II'', Compl.\ Univ.\ di Monte S. Angelo, 80126
Napoli, Italy}
\affiliation{Consorzio Nazionale Interuniversitario
per le Scienze Fisiche della Materia, Napoli}

\date{December 15, 2008}

\begin{abstract}
The orbital angular momentum of photons, being defined in an
infinitely dimensional discrete Hilbert space, offers a promising
resource for high-dimensional quantum information protocols in
quantum optics. The biggest obstacle to its wider use is presently
represented by the limited set of tools available for its control
and manipulation. Here, we introduce and test experimentally a
series of simple optical schemes for the coherent transfer of
quantum information from the polarization to the orbital angular
momentum of single photons and vice versa. All our schemes exploit
a newly developed optical device, the so-called ``q-plate'', which
enables the manipulation of the photon orbital angular momentum
driven by the polarization degree of freedom. By stacking several
q-plates in a suitable sequence, one can also access to
higher-order angular momentum subspaces. In particular, we
demonstrate the control of the orbital angular momentum $m$ degree
of freedom within the subspaces of $|m|=2\hbar$ and $|m|=4\hbar$
per photon. Our experiments prove that these schemes are reliable,
efficient and have a high fidelity.
\end{abstract}
 \maketitle

\section{Introduction}
Quantum information is based on the combination of classical
information theory and quantum mechanics. In the last few decades,
the development of this new field has opened far-reaching prospects
both for fundamental physics, such as the capability of a full
coherent control of quantum systems, as well as in technological
applications, most significantly in the communication field. In
particular, quantum optics has enabled the implementation of a
variety of quantum-information protocols. However, in this context,
the standard information encoding based on the two-dimensional
quantum space of photon polarizations (or ``spin'' angular momentum)
imposes significant limitations to the protocols that may be
implemented. To overcome such limitations, in recent years the
orbital angular momentum (OAM) of light, related to the photon's
spatial mode structure, has been recognized as a new powerful
resource for novel quantum information protocols, allowing the
implementation with a single photon of a higher-dimensional quantum
space, or a ``qu-dit'' \cite{Alle92,Moli07}. Thus far, the
generation of OAM-entangled photon pairs has been carried out by
exploiting the process of parametric down-conversion
\cite{Mair01,Moli03,Vazi03,Arna00,Frank02} and the quantum state tomography of such
entangled states has been achieved by using holographic masks \cite{Stuz07} and
single mode fibers \cite{Lang04}. The observation of pairs of
photons simultaneously entangled in polarization and OAM has been
also reported and exploited for quantum information protocols
\cite{Vazi02,Aiel05,Barr05,Barr08}.

Despite these successes, the optical tools for controlling the OAM
quantum states have been rather limited: the possibility of a wider
control and manipulation of the OAM resource, analogous to that
currently possible for the polarization degree of freedom, is yet to
be achieved.

As we demonstrate here, these limitations can be overcome by
exploiting the properties of an optical device, named ``q-plate'',
which has been recently introduced both in the classical and in the
quantum domain \cite{Marr06,Naga08}. The main feature of the q-plate
is its capability of coupling the spinorial (polarization) and
orbital contributions of the angular momentum of photons. The ease
of use, flexibility and good conversion efficiency showed by this
device makes it a promising tool for exploiting the OAM degree of
freedom of photons, in combination with polarization, as a resource
to implement high-dimensional quantum information protocols. In our
previous work \cite{Naga08} we provided a first demonstration of the
coherent information transfer capabilities for the simplest optical
schemes based on the q-plate, and for the case of a two-photon state
with non-classical correlations. In this paper we restrict our
attention to the single photon case and demonstrate, by quantum
tomography, the good coherence and fidelity of the information
transfer and the possibility to encode and then decode the
information within the orbital angular momentum space of a single
photon \cite{Gott08}. This last experiment is equivalent to a demonstration of
quantum communication taking place entirely in the orbital angular
momentum alphabet, with an overall efficiency that is in principle much higher
than previously demonstrated. Finally, we demonstrate the transfer of
quantum information to a higher order angular momentum subspace by
stacking more q-plates in sequence.

This paper is organized as follows. In Section II we illustrate the
q-plate device and its properties in the single photon regime. A
description of the different optical schemes adopted and of the
experimental setup is given in Section III. As mentioned above, we
have worked in the OAM SU(2) subspace $|m|=2$ (we will denote it as
$o_2$) and in the subspace $|m|=4$ (or $o_4$), where $m$ will
hereafter denote the quantum number giving the OAM per photon along
the beam axis in units of $\hbar$. Details on the hologram devices used
for the quantum tomography of photons in these OAM subspaces are
 given in Sec.\ IV. The demonstration of quantum information
transfer from the polarization quantum space to the OAM $o_2$
subspace and \emph{vice versa} is given in Sec.\ V, while Sec.\ VI
is concerned about the $o_4$ subspace. A brief conclusion is given
in Sec.\ VII.

\section{The q-plate}
A q-plate (QP) is a birefringent slab having a suitably patterned
transverse optical axis, with a topological singularity at its
center \cite{Marr06}. The ``charge'' of this singularity is given by
an integer or half-integer number $q$, which is determined by the
(fixed) pattern of the optical axis. The birefringent retardation
$\delta$ must instead be uniform across the device. Q-plates working
in the visible or near-infrared domain can be manufactured with
nematic liquid crystals, by means of a suitable treatment of the
containing substrates \cite{Marr06,Marr06b}. Once a liquid crystal
QP is assembled, the birefringent retardation $\delta$ can be tuned
either by mechanical compression (exploiting the elasticity of the
spacers that fix the thickness of the liquid crystal cell) or by
temperature control.

\begin{figure}[t!!]
\centering
\includegraphics[scale=.3]{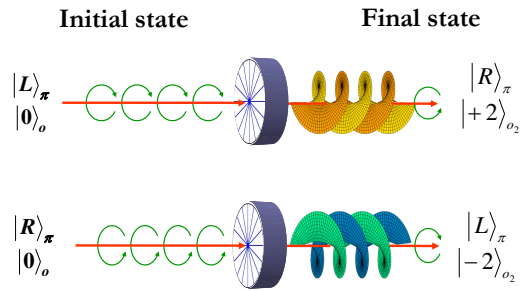}
\caption{Pictorial representation of the q-plate device action on a
photon quantum state.} \label{figqplate}
\end{figure}

For $\delta=\pi$, a QP modifies the OAM state $m$ of a light beam
crossing it, imposing a variation $\Delta{m}={\pm}2q$ whose sign
depends on the input polarization, positive for left-circular and
negative for right-circular \cite{Stal96}. The handedness of the output circular
polarization is also inverted, i.e. the optical spin is flipped \cite{Calv07}. In
the present work, we use only QPs with charge $q=1$ and
$\delta\simeq\pi$. Hence, an input TEM$_{00}$ mode (having $m=0$) is
converted into a beam with $m=\pm2$, as illustrated pictorially in
Fig.~\ref{figqplate}. In a single-photon quantum formalism, the QP
implements the following quantum transformations on the single
photon state:
\begin{eqnarray}
\ket{L}_{\pi}\ket{m}_{o} & \rightarrow & \ket{R}_{\pi}\ket{m+2}_{o} \nonumber\\
\ket{R}_{\pi}\ket{m}_{o} & \rightarrow &
\ket{L}_{\pi}\ket{m-2}_{o} \label{eqqplate}
\end{eqnarray}
where $\ket{\cdot}_{\pi}$ and $\ket{\cdot}_{o}$ stand for the
photon quantum state `kets' in the polarization and OAM degrees of
freedom,
 and $L$ and $R$ denote the left and right circular
polarization states, respectively. In the following, whenever there
is no risk of ambiguity, the subscripts $\pi$ and $o$ will be
omitted for brevity.

Any coherent superposition of the two input states given in
Eq. (\ref{eqqplate}) is expected to be preserved by the QP
transformation, leading to the equivalent superposition of the
corresponding output states \cite{Naga08}. Explicitly, we have
\begin{equation}
\alpha\ket{L}_{\pi}\ket{m}_{o}+\beta\ket{R}_{\pi}\ket{m}_{o}
\rightarrow
\alpha\ket{R}_{\pi}\ket{m+2}_{o}+\beta\ket{L}_{\pi}\ket{m-2}_{o}
\label{eqqplate2}
\end{equation}

Finally, we note that the quantum number $m$ does not completely
define the transverse mode of the photon. A radial number is also
necessary for spanning a complete basis, such as that of the
Laguerre-Gauss modes or an equivalent one. This radial degree of
freedom however does not play a significant role in the
demonstrations that will be reported in the following, because the
QP gives rise to a well defined radial profile, independent of the
sign of $m$ (see \cite{Kari08}), so for brevity in the
following we will omit it from our notations. This radial degree
of freedom may however become more critical when states having
different values of $|m|$ are manipulated simultaneously, a task which
will be addressed in a future paper. However we note that also holograms
create modes whose radial profiles are independent of the vortex sign \cite{Kari07}.
Such effect shows that our measurements are independent of radial index number.

\section{Experimental setup}
\begin{figure}[h]
\centering
\includegraphics[scale=.3]{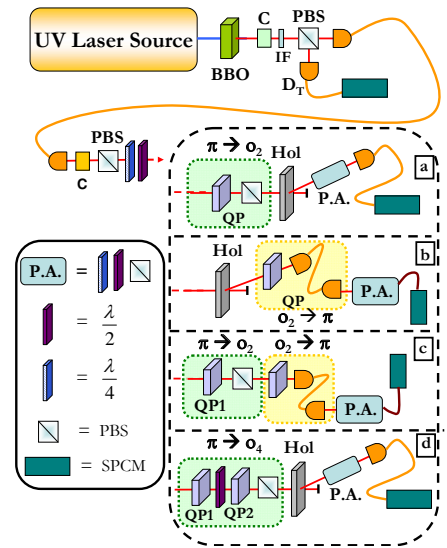}
\caption{Schematic representation of the experimental setup. Outside
the dashed box is the first section of the apparatus, common to all
our experiments. In the dashed box, the four configurations
\textbf{a, b, c, d} of the second section of the apparatus are
shown, used in the four experiments discussed in this paper. Legend:
BBO - $\beta$-barium borate crystal used for photon-pair generation;
C - walk-off compensation stage; PBS - polarizing beam-splitter;
D$_T$ trigger detection unit; QP - q-plate; Hol - hologram; S.P.C.M. - Single photon counter module; P.A. -
polarization analysis set, as shown in solid-line box.} \label{figsetup}
\end{figure}

Let us now describe the overall scheme of the experimental
apparatus, also shown in Fig.~\ref{figsetup}. The setup can be
divided in two main sections. The first one is common to all our
experiments and corresponds to the generator of triggered one-photon
states, with arbitrary polarization and fixed spatial mode
TEM$_{00}$. The second section is different for the four different
experiments (denoted as \textbf{a}, \textbf{b}, \textbf{c},
\textbf{d}) that will be described in the following Sections, and is
concerned with the OAM and polarization manipulations and with the
final quantum-state tomography.

In the first section of the apparatus, the main optical source is a
Ti:Sa mode-locked laser, with wavelength $\lambda=795$ nm, pulse
duration of 180 fs, and repetition rate 76 MHz. By second harmonic
generation, the output of the laser is converted into a ultraviolet
(UV) beam with wavelength $\lambda_p=397.5$ nm, which is used as
pump beam for the photon pairs generation. The residual field at
$\lambda$ is eliminated by means of a set of dichroic mirrors and
filters. The UV beam, with an average power of 600 mW, pumps a $1.5$
mm thick nonlinear crystal of $\beta$-barium borate (BBO) cut for
type II phase-matching, working in a collinear regime and generating
polarization pairs of photons with the same wavelength $\lambda$ and
orthogonal linear polarizations, hereafter denoted as horizontal
($H$) and vertical ($V$). These down-converted
photons are then spatially separated from the fundamental UV beam by
a dichroic mirror. The spatial and temporal walk-offs are
compensated by a half-wave plate and a 0.75 mm thick BBO (C)
\cite{Kwia95}. Finally, the photons are spectrally filtered by an
interference filter with bandwidth $\Delta\lambda=6$ nm.

In order to work in the one-photon regime, a polarizing
beam-splitter (PBS) transmits the horizontally-polarized photon of
the pair and reflects the vertically-polarized one. The latter is
then coupled to a single-mode fiber and revealed with a
single-photon counter (SPCM), which therefore acts as a trigger of
the one-photon state generation. The transmitted photon in the
$\ket{H}$ polarization state is coupled to another single-mode
fiber, which selects only a pure TEM$_{00}$ transverse mode,
corresponding to OAM $m=0$. The coincidence count rate of the two
outputs of the PBS, after coupling into the fibers, is of typically 16-18 kHz.

After the fiber output, two waveplates compensate (C) the
polarization rotation introduced by the fiber. Then, a polarizing
beam-splitter and a set of wave plates are
used for setting the photon polarization to an arbitrary qubit
state $\ket{\varphi}_{\pi}$. This concludes the first section of
the apparatus. The one-photon quantum state at this point can be
represented by the ket $\ket{\varphi}_{\pi}\ket{0}_{o}$.

Let us now consider the second main section of the apparatus. As we
mentioned above, this has been mounted in four different
configurations, shown in Fig.~\ref{figsetup}, corresponding to
the implementations of the following devices:\\
\textbf{a)} Quantum transferrer from polarization to OAM subspace $|m|=2$, i.e. $\pi\rightarrow o_2$\\
\textbf{b)} Quantum transferrer from OAM subspace $|m|=2$ to polarization, i.e. $o_2\rightarrow \pi$\\
\textbf{c)} Quantum bidirectional transfer polarization-OAM-polarization, i.e. $\pi\rightarrow o_2\rightarrow\pi$\\
\textbf{d)} Quantum transferrer from polarization to OAM subspace
$|m|=4$, i.e. $\pi\rightarrow o_4$\\

Each process of quantum information transfer is based on a q-plate (two in
the cases c and d) combined with other standard polarizers and
waveplates. The OAM state is prepared or analyzed by means of
suitably-developed holograms, as discussed in the next Section,
preceded or followed by coupling to single-mode fibers, which
selects the $m=0$ state $\ket{0}_{o}$ before detection. After
the analysis, the signals have been detected by single photon
counters SPCM and then sent to a coincidence box interfaced with a
computer, for detecting and counting the coincidences of the
photons and the trigger D$_T$.

\section{Holograms and OAM-polarization correspondence}
A full analogy can be drawn between the polarization SU(2) Hilbert
space and each subspace of OAM with a given $|m|$, except of course
for $m=0$. This analogy is for example useful for retracing the
quantum tomography procedure to the standard one for polarization
\cite{Padg99,Lang04}. In particular, it is convenient to consider
the eigenstates of OAM $\ket{\pm|m|}$ as the analog of the circular
polarizations $\ket{L}$ and $\ket{R}$, as the latter ones are
obviously the eigenstates of the spin angular momentum. To make the
analogy more apparent, small-letter symbols
$\ket{l}=\ket{+|m|}$ and $\ket{r}=\ket{-|m|}$ are introduced to
refer to the OAM case, while the capital letters are used for the
polarization. Following the same convention, the OAM equivalent of
the two basis linear polarizations $\ket{H}$ and $\ket{V}$ are then
defined as
\begin{eqnarray}
\ket{h}&=&\frac{1}{\sqrt{2}}(\ket{l}+\ket{r})\nonumber\\
\ket{v}&=&\frac{1}{i\sqrt{2}}(\ket{l}-\ket{r})
\end{eqnarray}
Finally, the $\pm45^{\circ}$ angle ``anti-diagonal'' and
``diagonal'' linear polarizations will be hereafter denoted with the
kets $\ket{A}=(\ket{H}+\ket{V})/\sqrt{2}$ and
$\ket{D}=(\ket{H}-\ket{V})/\sqrt{2}$, and the corresponding OAM
states are defined analogously:
\begin{eqnarray}
\ket{a}&=&\frac{1}{\sqrt{2}}(\ket{h}+\ket{v})=\frac{e^{-i\pi/4}}{\sqrt{2}}(\ket{l}+i\ket{r})\nonumber\\
\ket{d}&=&\frac{1}{\sqrt{2}}(\ket{h}-\ket{v})=\frac{e^{i\pi/4}}{\sqrt{2}}(\ket{l}-i\ket{r}).
\end{eqnarray}

\begin{figure}[h]
\centering
\includegraphics[scale=0.35]{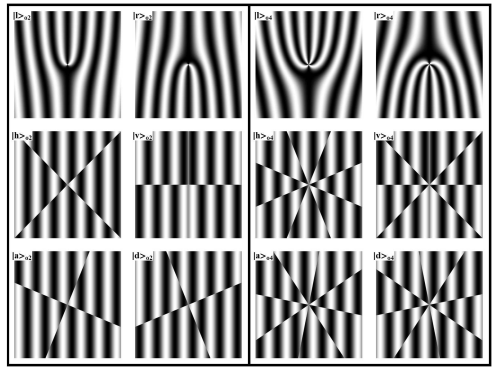}
\caption{Patterns of the 12 binary holograms used in this work.
The left box refers to the OAM subspace $o_2$ ($|m|=2$). The right
box to the OAM subspace $o_4$ ($|m|=4$). In the upper-left corner
of each hologram is shown the quantum state that is generated by
the hologram, when using a TEM$_{00}$ input, in the first-order
diffraction beam.} \label{fighol}
\end{figure}

The holograms used for generating or analyzing the above OAM states
were manufactured from a computer-generated image by a photographic
technique followed by a chemical bleaching step, producing
pure phase binary holographic optical elements. The typical
first-order diffraction efficiencies of these holograms were in the
range 10-15\%. The patterns we used are shown in Fig.~\ref{fighol}.
Analogously to polarizers, these holograms are used in two ways: (i)
for generating a given input quantum state of OAM; (ii) for
analyzing a given OAM component of an arbitrary input quantum state.

When using the holograms for generating one of the above OAM states,
a TEM$_{00}$ input mode is sent into the hologram and the
first-order diffracted mode is used for output. The input beam must
be precisely centered on the hologram pattern center. The output OAM
quantum state obtained is shown in the upper corner of each hologram
pattern in Fig.~\ref{fighol}.

When using the holograms for analysis, the input mode, having
unknown OAM quantum state, is sent through the hologram (with proper
centering). The first-order diffracted output is then coupled to a
single-mode fiber, which filters only the $m=0$ state, before
detection. It can be shown that the amplitude of this output is then
just proportional to the projection of the input state onto the OAM
state shown in the upper corner of each hologram pattern, in
Fig.~\ref{fighol} (except, possibly, for a sign inversion of $m$ in
the case of the upper row holograms).
\section{Manipulation of orbital angular momentum in the subspace $|m|=2$}
A single q-plate (with $q=1$) can be used for coupling the
polarization subspace $\pi$ with the OAM subspace $o_2$, spanned by the OAM
eigenstates $\{\ket{+2}_{o},\ket{-2}_{o}\}$. In this Section,
we present two optical schemes based on the q-plate that enable a
qubit of quantum information to be transferred from the
polarization to the OAM (setup \textbf{a}, \emph{transferrer}
$\pi\rightarrow o_2$), from OAM to polarization (setup \textbf{b},
\emph{transferrer} $o_2\rightarrow\pi$). Moreover, we tested also
the combination of these two schemes, thus realizing the
\emph{bidirectional transfer} polarization-OAM-polarization (setup
\textbf{c}, $\pi\rightarrow o_2\rightarrow\pi$). The latter
demonstration is equivalent to demonstrating quantum communication
using OAM for encoding the message. In other words, the qubit is
initially prepared in the polarization space, then passed to OAM
in a transmitting unit (Alice), sent to a receiving unit (Bob),
where it is transferred back to polarization for further
processing or detection.

All these transfer processes have been experimentally verified by
carrying out quantum tomography measurements, either in the
polarization or in the OAM degree of freedom. The latter was based
on the polarization - OAM subspace analogy discussed in the previous
Section. Let us now see the details of each of the three schemes.

\subsection{Transferrer polarization to OAM}
Let us consider as initial state the polarization-encoded qubit
\begin{equation}
\ket{\Psi}_{\text{in}}=\ket{\varphi}_{\pi}\ket{0}_{o}=(\alpha\ket{H}_{\pi}+\beta\ket{V}_{\pi})\ket{0}_{o}
\end{equation}
where $\ket{0}_{o}$ indicates the TEM$_{00}$ mode. By passing it
through a pair of suitably oriented quarter-waveplates (one with
the optical axis parallel to the horizontal direction and the
other at 45$^{\circ}$), the photon state is rotated into the $L,R$
basis:
\begin{equation}
(\alpha\ket{L}_{\pi}+\beta\ket{R}_{\pi})\ket{0}_{o}
\end{equation}
After the QP the quantum state of the photon is then turned into the
following:
\begin{equation}
\alpha\ket{R}\ket{+2}+\beta\ket{L}\ket{-2}.
\end{equation}
If a polarizer along the horizontal direction is used, we then
obtain the state
\begin{equation}
\ket{\Psi}_{out}=\ket{H}_{\pi}(\alpha\ket{+2}_{o}+\beta\ket{-2}_{o})=\ket{H}_{\pi}\ket{\varphi}_{o_2},
\end{equation}
which completes the conversion. We note that such conversion
process is probabilistic, since the state $\ket{\Psi}_{out}$ is
obtained with a probability $p=50\%$, owing to the final
polarizing step. Moreover, since we are using the
$\{\ket{H},\ket{V}\}$ basis for the polarization encoding and the
$\{\ket{+2},\ket{-2}\}=\{\ket{l},\ket{r}\}$ for the OAM one, the
transfer is associated also with a rotation of the Poincar\'e
sphere. The correspondence of the six orthogonal states on the
polarization Poincar\'e sphere with the six final ones in the OAM
sphere is given in Table 1.
\begin{figure}[h]
\centering
\includegraphics[scale=.3]{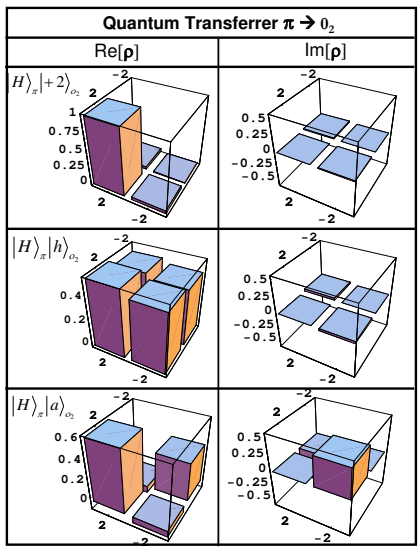}
\caption{Experimental density matrices $\rho$ (the left column shows
the real part and right column the imaginary part) measured for the
output of the $\pi\rightarrow o_2$ qubit transfer, for each of the
three different predicted output states shown in the upper left
corner of each row.} \label{fig_pi_o2}
\end{figure}

The experimental layout of this scheme is shown in
Fig.~\ref{figsetup}, dashed box \textbf{a}. The input arbitrary
qubit is written in the polarization using two waveplates, as
discussed in Sec.\ III. The final state tomography has been
realized by means of the six holograms shown in Fig.~\ref{fighol}
(left box). The experimental results for three specific choices of
the input state are shown in Fig.~\ref{fig_pi_o2}. We find a good
agreement with theory as demonstrated by the fidelity parameter,
defined as $F=\bra{\psi}\rho_{exp}\ket{\psi}$, where $\ket{\psi}$
is the theoretical state to be compared to the experimental one.
Hence in this experiment the average fidelity value between the
experimental states and the theoretical predictions is
$F=(97.7\pm0.2)\%$. The fidelities obtained for six different
input states are shown in Table 1.
\begin{table}
\begin{center}
\begin{tabular}{||c|c|c||}
\hline\hline
\textbf{Initial state} & \textbf{Final state} & \textbf{Fidelity}\\
\hline\hline
$\ket{H}_{\pi}$ & $\ket{+2}=\ket{l}_{o_2}$ & $(0.990\pm0.002)$ \\
\hline
$\ket{V}_{\pi}$ & $\ket{-2}=\ket{r}_{o_2}$ & $(0.972\pm0.002)$ \\
\hline\hline
$\ket{A}_{\pi}$ & $\ket{h}_{o_2}$ & $(0.981\pm0.002)$ \\
\hline
$\ket{D}_{\pi}$ & $\ket{v}_{o_2}$ & $(0.968\pm0.002)$ \\
\hline\hline
$\ket{L}_{\pi}$ & $\ket{a}_{o_2}$ & $(0.998\pm0.002)$ \\
\hline
$\ket{R}_{\pi}$ & $\ket{d}_{o_2}$ & $(0.982\pm0.002)$ \\
\hline\hline
\end{tabular}
\end{center}
\caption{Fidelity values between the experimental states generated
by the $\pi\rightarrow o_2$ transferrer and the theoretical ones
expected after the conversion in the OAM degree of freedom of the
qubit initially encoded in the polarization.}
\end{table}

Thus, we have demonstrated experimentally that the initial
information encoded in an input TEM$_{00}$ state can be coherently
transferred to the OAM degree of freedom, thanks to the
$\pi\rightarrow o_2$ converter, giving rise to the preparation of a
qubit in the orbital angular momentum. As the initial information
has been stored in the orbital part of the qubit wave-function, new
information can be stored in the polarization degree of freedom,
allowing the transportation in a single photon of a higher amount, at least
two qubits, of
information.

\subsection{Transferrer OAM to polarization}
Let us now show that the reverse process can be realized as well, by
transferring a qubit initially encoded in the OAM subspace $o_2$
into the polarization space. We therefore consider as initial
quantum state of the photon the following one:
\begin{equation}
\ket{\Psi}_{\text{in}}=\ket{H}_{\pi}\ket{\varphi}_{o_2}=\ket{H}(\alpha\ket{+2}+\beta\ket{-2})
\end{equation}
By injecting the state $\ket{\Psi}_{in}$ in the q-plate device, and
then rotating the output state by means of a pair of waveplates, we
obtain the following state:
\begin{equation}
\frac{1}{2}\{\alpha\ket{V}\ket{+4}+\alpha\ket{H}\ket{0}+\beta\ket{V}\ket{0}+\beta\ket{H}\ket{-4}\}
\end{equation}
Now, by coupling the beam to a single mode fiber, only the states
with $m=0$ that is, the TEM$_{00}$ modes, will be efficiently
transmitted. Of course, this implies that a probabilistic process is
obtained again, since we discard all the contributions with
$m\neq0$ (ideally, again $p=50\%$). After the fiber, the output
state reads:
\begin{equation}
\ket{\Psi}_{out}=(\alpha\ket{H}+\beta\ket{V})\ket{0}=\ket{\varphi}_{\pi}\ket{0}_{o}
\end{equation}
which demonstrates the successful conversion from the OAM degree
of freedom to the polarization one.

\begin{figure}[h]
\centering
\includegraphics[scale=.3]{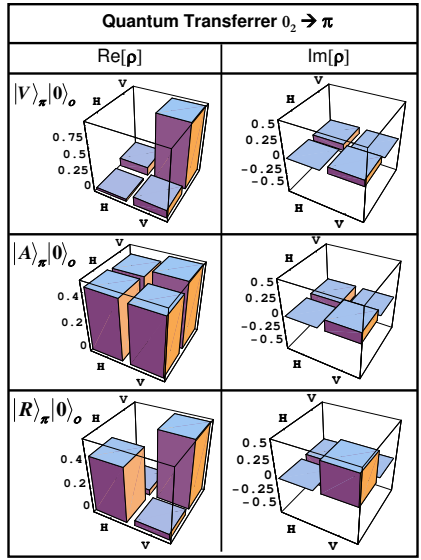}
\caption{Experimental density matrices $\rho$ (the left column shows
the real part and right column the imaginary part) measured for the
output of the $o_2\rightarrow \pi$ qubit transfer, for each of the
three different predicted output states shown in the upper left
corner of each row.} \label{fig_o2_pi}
\end{figure}

The experimental layout of this ``reverse'' scheme is shown in
Fig.~\ref{figsetup}, dashed box \textbf{b}. The input qubit in OAM
is prepared using one of the six holograms shown in
Fig.~\ref{fighol} (left box), as explained in the previous Section.
The output state is analyzed by a standard polarization-state
quantum tomography. The experimental results for three cases are
shown in Fig.~\ref{fig_o2_pi}. We find again a good agreement with
theory, with an average fidelity $F=(97.3\pm0.2)\%$, and the
specific cases shown in Table 2.

\begin{table}
\begin{center}
\begin{tabular}{||c|c|c||}
\hline\hline
\textbf{Initial state} & \textbf{Final state} & \textbf{Fidelity}\\
\hline\hline
$\ket{+2}=\ket{l}_{o_2}$ & $\ket{H}_{\pi}$ & $(0.981\pm0.002)$ \\
\hline
$\ket{-2}=\ket{r}_{o_2}$ & $\ket{V}_{\pi}$ & $(0.995\pm0.002)$ \\
\hline\hline
$\ket{a}_{o_2}$ & $\ket{L}_{\pi}$ & $(0.964\pm0.002)$ \\
\hline
$\ket{d}_{o_2}$ & $\ket{R}_{\pi}$ & $(0.972\pm0.002)$ \\
\hline\hline
$\ket{h}_{o_2}$ & $\ket{A}_{\pi}$ & $(0.967\pm0.002)$ \\
\hline
$\ket{v}_{o_2}$ & $\ket{D}_{\pi}$ & $(0.970\pm0.002)$ \\
\hline\hline
\end{tabular}
\end{center}
\caption{Fidelity values between the experimental states generated
by the $o_2\rightarrow\pi$ transferrer and the theoretical ones
expected after the conversion in polarization degree of freedom of
the qubit initially encoded in the OAM.}
\end{table}

We note that this OAM-to-polarization transferrer allows a simple
detection of the sign of the OAM, with a theoretical efficiency of
$50\%$, much larger than what is typically obtained by the fork holograms
($10\%\div30\%$). Therefore, this scheme can be used as a very efficient
OAM detector.

\subsection{Bidirectional transfer polarization-OAM-polarization}
Having demonstrated polarization-to-OAM transfer and
OAM-to-polarization transfer, it is natural to try both schemes
together, in a bidirectional transfer which starts and ends with
polarization encoding, with OAM as an intermediate state which can
be used for example for communication. This is also the first
quantum experiment based on the combined use of two q-plates.
Although this test in principle is not involving any new idea with
respect to the previous two experiments, it is important to verify
that in practice the efficiency of the optical manipulation is not
strongly affected by the number of q-plate employed, for example due
to alignment criticality.

\begin{figure}[h]
\centering
\includegraphics[scale=.3]{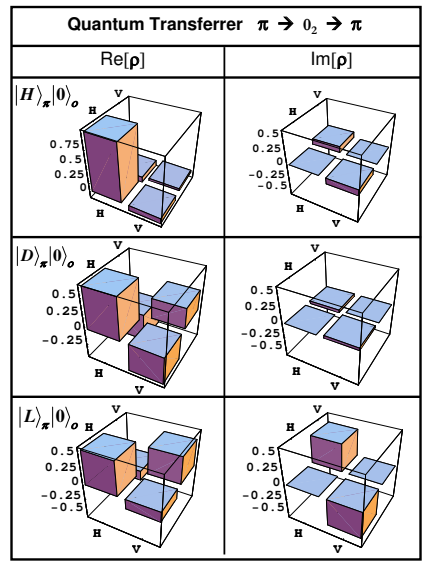}
\caption{Experimental density matrices measured in the polarization
degree of freedom after the bidirectional $\pi\rightarrow
o_2\rightarrow \pi$ transferrer. In each box is reported the
expression of the initial and final state, to be compared with the
experimental one described by the density matrix.} \label{fig_pi_o2_pi}
\end{figure}

The layout is shown in Fig.~\ref{figsetup}, dashed box \textbf{c},
and corresponds to the sequence of the two schemes discussed above.
In Fig.~\ref{fig_pi_o2_pi} we show some density matrices obtained by
the quantum tomography technique in the polarization degree of
freedom of the output state.

As can be observed in Table 3, the experimental results are in good
agreement with the theoretical predictions, with a mean fidelity
value equal to $F=(95.9\pm0.2)\%$. We see that the overall fidelity
is still quite good, so that there seem to be no significant problem
to the combined use of many q-plates in a cascaded configuration. After the
two q-plates the quantum efficiency of the conversion process, defined as
the capability to convert a TEM$_{00}$ mode in a pure Laguerre-Gauss, is still around $80\%$ (to optimize the
efficiency, the q-plate birefringent retardations $\delta$ were
tuned by mechanical pressure).
\begin{table}
\begin{center}
\begin{tabular}{||c|c||}
\hline\hline
\textbf{Initial and final state} & \textbf{Fidelity}\\
\hline\hline
$\ket{H}_{\pi}$ & $(0.970\pm0.002)$ \\
\hline
$\ket{V}_{\pi}$ & $(0.972\pm0.002)$ \\
\hline\hline
$\ket{A}_{\pi}$ & $(0.958\pm0.002)$ \\
\hline
$\ket{D}_{\pi}$ & $(0.955\pm0.002)$ \\
\hline\hline
$\ket{R}_{\pi}$ & $(0.934\pm0.002)$ \\
\hline
$\ket{L}_{\pi}$ & $(0.962\pm0.002)$ \\
\hline\hline
\end{tabular}
\end{center}
\caption{Fidelity values between the input and output states for the
bidirectional $\pi\rightarrow o_2\rightarrow \pi$ transferrer.}
\end{table}

\subsection{Deterministic conversion processes}
\begin{figure}[h]
\centering
\includegraphics[scale=.3]{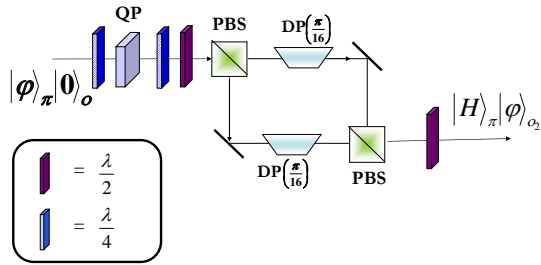}
\caption{Schematic representation of deterministic transferrer: A pair
of suitably rotated Dove's prisms (DP), combined with
wave plates for polarization compensation, are used to realize a $sigma_z$
operation on the OAM degree of freedom. The transferrer converts
the polarization in OAM if the light goes from left to right, while
it converts the OAM into polarization if crossed from right to left.} \label{figdettransf}
\end{figure}

The quantum transferrers implemented experimentally up to now are
probabilistic processes, with 50\% success probability. However, we
now show that it is possible to realize a fully
\emph{deterministic} transferrer for both directions
polarization-OAM and backward. This is obtained at the price of a
slightly more complex optical layout, based on a q-plate and a
Mach-Zehnder interferometer, shown in Fig.~\ref{figdettransf}. The
deterministic transferrer is bidirectional, and it converts the
polarization in OAM ($\pi\rightarrow o_2$) if crossed in one way and
the OAM in polarization ($o_2\rightarrow \pi$) if crossed in the
opposite way.

Let us consider first the $\pi\rightarrow o_2$ conversion. The
initial state reads:
\begin{equation}
\ket{\Psi}_{in}=\ket{\varphi}_{\pi}\ket{0}_{o}=(\alpha\ket{H}+\beta\ket{V})\ket{0}
\end{equation}
A pair of quarter waveplates converts it into the $L,R$ basis, and
then the QP is applied, so as to obtain the following state:
\begin{equation}
\alpha\ket{R}\ket{+2}+\beta\ket{L}\ket{-2}
\end{equation}
Another set of half-wave plates rotate the polarization basis in
${\ket{A},\ket{D}}$, leading to $\alpha\ket{A}\ket{+2}+\beta\ket{D}\ket{-2}$:
\begin{eqnarray}
\frac{1}{\sqrt{2}}\left(\ket{H}(\alpha\ket{+2}+\beta\ket{-2})+\ket{V}(\alpha\ket{+2}
-\beta\ket{-2})\right)
\label{eqdettrasf}
\end{eqnarray}

Such state is then injected in a PBS that separates the two linear
polarizations and sends them in the two arms of a Mach-Zehnder
interferometer. In one arm of the interferometer, say the
$V$-polarized one, a device acting as a Pauli's operator
$\tilde{\sigma}_z$ is inserted that operates only on the OAM states.
This operator can be for example realized by means of a Dove's prism
rotated at a $\pi/8$ angle in the lab frame followed by another
Dove's prism rotated at zero angle, eventually with a set of
compensating wave-plates for correcting possible polarization
variations. Alternatively, one Dove's prism can be put in one arm
and the other in the other arm of the interferometer (to make it
more balanced), both rotated by $\pi/16$. At each reflection in a
mirror or in the PBS (as well as in a Dove's prism) the OAM is
flipped ($m\rightarrow -m$). However, the overall number of
reflections is even in both paths, so we can ignore this effect
(however, some care must be taken for computing the correct phases
of each term).

Mathematically, the $\tilde{\sigma}_z$ device will just change sign
to the last term in Eq.~(\ref{eqdettrasf}). Therefore, the state in
the interferometer becomes the following:
\begin{equation}
\ket{H}\frac{1}{\sqrt{2}}(\alpha\ket{+2}+\beta\ket{-2})+\ket{V}\frac{1}{\sqrt{2}}(\alpha\ket{+2}
+\beta\ket{-2}) \label{eqdettrasf}
\end{equation}
where it is understood that $\ket{H}$ is also associated with one
arm and $\ket{V}$ with the other arm of the interferometer. After
the exit PBS, these two states are again superimposed in the same
mode and provide only a single output on one exit
face of the PBS, which is the following:
\begin{equation}
\ket{A}(\alpha\ket{+2}+\beta\ket{-2})
\end{equation}
The polarization state is then finally rotated to $H$ by a final
half-wave plate rotated by 22.5$^{\circ}$. Thus, the expected final
state
\begin{equation}
\ket{\Psi}_{out}=\ket{H}(\alpha\ket{+2}+\beta\ket{-2})=\ket{H}_{\pi}\ket{\varphi}_{o_2}
\end{equation}
is obtained, this time deterministically, as no contribution has
been discarded \cite{nota}.
The opposite conversion, $o_2\rightarrow \pi$, is obtained by simply
reversing the direction of light propagation in the same setup. All
the transformations are then reversed and provide the desired
information transfer from OAM to polarization, again fully
deterministically.

\section{Manipulation of orbital angular momentum in the subspace $|m|=4$}
In the bidirectional transfer, we have experimentally demonstrated
that it is possible to work with two sequential q-plates without a
significant lowering of the overall efficiency. This approach can be
also adopted to access higher-order subspaces of the orbital angular
momentum, by moving from one subspace to the next using a sequence
of QPs alternated with half-wave plates \cite{Marr06b}.

Experimentally we have studied the case of two sequential q-plates
QP$_1$ and QP$_2$ (both with $q=1$). We demonstrate that it is
possible to efficiently encode the quantum information in the OAM
basis $\{\ket{+4},\ket{-4}\}$, by exploiting the spin-orbit coupling
in the q-plates. In order to analyze the orbital angular momentum
with $|m|=4$ we have adopted newly designed holograms, shown in
Fig.~\ref{fighol} (box on the right).

An initial state in the TEM$_{00}$ mode and arbitrary polarization
$\ket{\varphi}_{\pi}=(\alpha\ket{H}+\beta\ket{V})$ is transformed by
a pair of quarter-wave plates and QP$_1$ into the following one:
\begin{equation}
\ket{\varphi}_{\pi}\ket{0}_l\rightarrow(\alpha\ket{R}\ket{-2}+\beta\ket{L}\ket{+2})
\end{equation}
\begin{figure}[h]
\centering
\includegraphics[scale=.32]{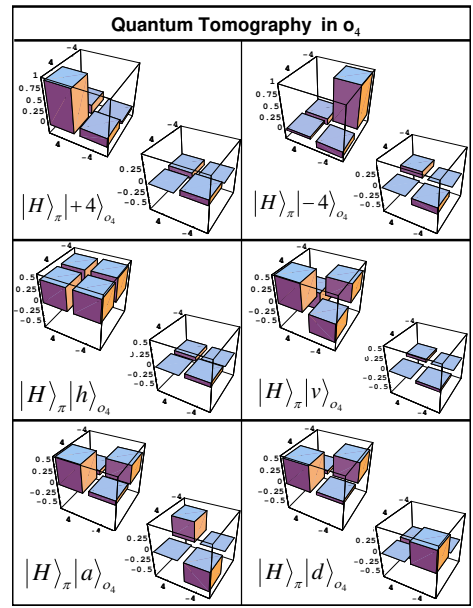}
\caption{Experimental density matrices measured in the OAM basis
$\{\ket{+4},\ket{-4}\}$ for different predicted final states, shown
in the lower-left corner of each panel.} \label{fig_pi_o4}
\end{figure}
A half-wave plate then inverts the polarization of the output state
after QP$_1$, so that we get:
\begin{equation}
\alpha\ket{L}\ket{+2}+\beta\ket{R}\ket{-2}
\end{equation}
Next, the action of QP$_2$ and a polarizer leads to the final state:
\begin{equation}
(\alpha\ket{+4}+\beta\ket{-4})\ket{H}=\ket{\varphi}_{o_4}\ket{H}_{\pi}
\end{equation}

By changing the different hologram masks, we have carried out the
quantum state tomography reported in Fig.~\ref{fig_pi_o4}. The
fidelity related to each state is reported in Table 4, and the high
accordance between theory and experimental data leads to an average
value $F=(96.1\pm0.2)\%$.
\begin{table}
\begin{center}
\begin{tabular}{||c|c|c||}
\hline\hline
\textbf{Initial state} & \textbf{Final state} & \textbf{Fidelity}\\
\hline\hline
$\ket{H}_{\pi}$ & $\ket{+4}=\ket{l}_{o_4}$ & $(0.947\pm0.002)$ \\
\hline
$\ket{V}_{\pi}$ & $\ket{-4}=\ket{r}_{o_4}$ & $(0.958\pm0.002)$ \\
\hline\hline
$\ket{L}_{\pi}$ & $\ket{a}_{o_4}$ & $(0.992\pm0.002)$ \\
\hline
$\ket{R}_{\pi}$ & $\ket{d}_{o_4}$ & $(0.923\pm0.002)$ \\
\hline\hline
$\ket{A}_{\pi}$ & $\ket{h}_{o_4}$ & $(0.994\pm0.002)$ \\
\hline
$\ket{D}_{\pi}$ & $\ket{v}_{o_4}$ & $(0.955\pm0.002)$ \\
\hline\hline
\end{tabular}
\end{center}
\caption{Fidelity values between the expected and the experimental
states generated by the $\pi\rightarrow o_4$ transferrer.}
\end{table}

\section{Conclusion}
In this work we have demonstrated several optical schemes for
the efficient quantum manipulation of the orbital angular
momentum degree of freedom of single photons. All these schemes
are based on the q-plate, a novel optical device that introduces a
coupling between the polarization and the orbital angular
momentum. We have experimentally demonstrated the coherent
transfer of a qubit from the polarization to the orbital angular
momentum and vice versa. We have also demonstrated the scalability
of this approach, by cascading two q-plates in order to accomplish
(i) the bidirectional transfer from the polarization to the
orbital angular momentum and back to polarization and (ii) access
to higher orders of orbital angular momentum. In all these
demonstrations we achieved very good fidelities, as calculated by
quantum tomographies of the resulting qubits, and also good
quantum efficiencies. The schemes demonstrated experimentally are
probabilistic, with 50\% success rate. However, we have also
proposed a related scheme that is fully deterministic, although
slightly more complex.

These results simplify the use of orbital angular momentum for encoding
the quantum information in a single photon and can make its manipulation more
convenient, by linking this orbital degree of freedom to the standard
one of polarization. In perspective, this approach would
allow realizing simple and effective schemes for higher dimensional
quantum information processing and communication with photons.



\begin{thebibliography}{99}
\bibitem{nota} We note that, although the optical layout is a Mach-Zehnder interferometer, the optical path phase difference between the two arms of the interferometer is only affecting the polarization state of the single output obtained after the final PBS, while it does not act on the PBS exit mode and on the OAM final state. The final polarization may therefore turn elliptical if this phase difference is not well controlled. However, the H polarization can be easily restored by suitable wave-plates, as long as it is uniform. Therefore, this setup is expected to be relatively robust against misalignment.
\end{thebibliography}
\end{document}